\documentstyle[prd,aps,epsfig]{revtex}
\begin{document}
\draft
\author{Yuriy Mishchenko and Chueng-Ryong Ji}
\address{Department of Physics, North Carolina State University,
Raleigh, North Carolina 27695-8202}
\title{Distribution of mass in galaxy cluster CL0024 and
the particle mass of dark matter.}
\date{\today}
\maketitle

\begin{abstract}
We study in details the distribution of mass in galaxy
cluster CL0024+1654 inferred using the method of strong
gravitational lensing by Tyson {\it et al.} (1998).
We show that a linear correlation exists between total,
visible and dark matter distributions on log-log scale
with consistent coefficients. The shape and parameters
of log-log-linear correlation are not affected significantly
whether one uses projected or volume mass densities
but is consistent with $\kappa=2-5$ visible/dark ratio. We also
show and analyze in depth so called alignment properties of the
above-mentioned profiles. We show that log-log-linear
correlation and alignments can all be understood in
terms of thermodynamic/hydrodynamic equilibrium with
gravitational potential growing almost linearly in the
region of interest.
We then analyze
the hypothesis of thermal equilibrium on 
the base of the existing data about
CL0024 cluster.
If the presence of log-log-linear correlation and alignments
were interpreted thermodynamically, this would
indicate the mass of the dark matter particle 2-5 times
smaller than that of atomic hydrogen, thus giving range for
the mass of dark matter particle between 200MeV and 1000MeV.
\end{abstract}

\section{Introduction}
In the past decade the field of astronomy and astrophysics has experienced
a period of rapid development thanks to the operation of 
Hubble Space Telescope and a wide spread of new faster computers. 
This development brought a series of brilliant
insights in the puzzles of the structure and history of the universe.
At this time the growing number
of precision measurements of galactic rotation curves and
observations of gravitational lensing in galaxies and galaxy clusters
combined with extensive computer simulations shed new light
on the problem of hidden, or dark, matter in the universe
and its role in the universe evolution.
Still, little is 
known about the microscopic properties and composition of the dark matter. 
Many possibilities have been 
 put forward by various extensions of the
Standard Model but yet existence of any of 
such particles is to be confirmed 
experimentally.

Recently it had been pointed out that the distribution of dark and
visible matter in spiral galaxies with various luminosities and galaxy clusters 
exhibit strikingly similar correlations with almost identical parameters \cite{ji}. 
Specifically, it was found that 
in all of these systems
the mass densities of visible and dark matter
are linearly correlated on log-log scale.
The proportionality coefficient in this correlation
appears to be {\em universal} and equal to $\kappa\sim 3-5$.
This conclusion had been drawn from the analysis of dark and
visible matter distribution in the galaxy cluster CL0024 \cite{tyson} 
and synthetic mass model for 
spiral galaxy  rotation curves of
Persic and Salucci \cite{persic}.

In this paper we are going to further sharpen our focus 
on the properties of the mass distribution in
galaxy cluster CL0024. 
We will investigate in details presence and properties of 
log-log-linear correlations, alignment properties and their 
thermodynamic significance 
and interpretation.

We begin in the next section (Section \ref{secII}) with 
discussion of the properties of the projected mass density 
profiles $\Sigma(R)$ derived 
by Tyson {\it et al.} in 1998 and focus on the log-log-linear correlation
between them. 
We then continue with analysis of the volume density
profiles $\rho(R)$ inferred from $\Sigma(R)$ via inverse Abel transformation
and show that the log-log-linear correlation is present also
between the volume density profiles.
We comment on the significance of these correlations
and their thermodynamic interpretation.
Finally,  we review the alignment properties of the mass profiles
in the galaxy cluster CL0024, mentioned earlier in the literature, and show
that they imply exponential behavior for the radial mass density $\Sigma(R)$ in
the region of interest.
In Section \ref{secIII} we present  analysis of the thermal state
of the matter in the galaxy cluster based on the strong
gravitational lensing study carried out by Tyson {\it et al.}
Summary and conclusions follow in Section \ref{conclusions}.

\section{Mass distribution in the galaxy cluster CL0024}
\label{secII}
That 
the mass distribution in galaxies or galaxy clusters can be 
measured using gravitational lensing
had been pointed out long time ago 
\cite{straumann}. 
Gravitational lensing is one of the consequences of Einstein's 
theory of gravity (General Relativity) in which the light from a distant object 
is bent by a heavy galaxy or galaxy cluster
to produce "fake" images of the original object.
The amount of this distortion is very tiny and for an observable effect
to be seen the objects' alignment should be quite perfect.
Still, with billions of galaxies in the sky, a number of gravitational lenses
had been discovered in recent years providing excellent
tests of the Theory of General Relativity.
One of the most famous examples of gravitational lensing
is the object known as Einstein's cross (or Q2237+030), in which the
light from a distant quasar is disrupted by a low-redshift galaxy
to produce four images
in the form of a cross surrounding the galaxy's nucleus.

Galaxy clusters are one of the most interesting objects for
gravitational lensing studies because of their unsurpassed mass
and large extent. In fact, precision measurement of gravitational
lensing in a galaxy cluster is capable to produce a detailed map 
of gravitating mass distribution inside the cluster and
thus provide valuable information
about its structure.
In the last decade many clusters exhibiting
gravitational lensing have been found
and their "gravitating" mass distributions have been analyzed
\cite{tyson,wu,kneib,abdelsalam,bezecourt,white}.
It was generally shown that a large portion of mass in galaxy clusters
is not associated with the luminous galaxies and form a smooth extended distribution.
However, practically never such mass maps, obtained from
lensing, have been compared with the 
distribution of the
visible matter inside the corresponding clusters.
The study of strong gravitational lensing in galaxy cluster
CL0024 by Tyson {\it et al.} in 
1998 \cite{tyson} is distinguished in that the detailed maps 
both for total and visible mass in the cluster were constructed and presented.

The galaxy cluster CL0024+1654 is a remarkable instance of
strong gravitational lensing in which multiple images of
a background galaxy with distinctive spectrum are formed.
In 1998, an analysis of Hubble telescope images of this cluster
was carried out by Tyson {\it et al.} and the mass
profile of the cluster was obtained from strong gravitational 
lensing.
Tyson {\it et al.} found that the vast majority of the mass in CL0024
is not associated
with the galaxies and forms a smooth elliptical distribution,
slightly shallower than isothermal sphere, with a soft core 
of $r_{core}=35\pm3 h^{-1} kpc$, where $h$ is the normalized Hubble
constant. No evidence of in-falling massive clumps had been found 
for the dark component.
The projected dark matter density profile was fit well by a power-law model
\begin{equation}
\Sigma(y)=\frac{K(1+\eta y^2)}{(1+y^2)^{2-\eta}},
\end{equation}
where $y=r/r_{core}$, $K=7900\pm100 h M_\odot pc^{-2}$, 
$r_{core}=35\pm3 h^{-1}kpc$ and $\eta=0.57\pm0.02$. 
The primary conclusion was 
that the cusped mass profile for the dark component,  suggested by 
many-body simulations within the
Cold Dark Matter model
\cite{navaro}, was 
inconsistent with the observed results.
Along with the total mass distribution
Tyson {\it et al.} also presented the radial distribution of the visible matter density
and of the visible light density.

It was recently noticed that the mass profiles obtained in Ref.\cite{tyson}
possess interesting correlation properties \cite{ji}. Specifically, if the mass profile
of the dark matter and that of the visible matter are plotted one vs. the other
on log-log scale, the linear correlation between the two becomes apparent which
implies
\begin{equation}
\log \Sigma_v \approx \kappa_{vd} \log \Sigma_d.
\end{equation}

Remarkably, such behavior of mass profiles should be expected
on thermodynamical grounds.
For example, for isothermal distribution of self-gravitating
gas it is known that the mass profiles satisfy
\begin{equation}\label{equi}
\begin{array}{c}
\log \rho_v \sim e^{-\frac{\mu_v \Phi(r)}{T}}, \\
\log \rho_d \sim e^{-\frac{\mu_d \Phi(r)}{T}},
\end{array}
\end{equation}
where $\mu_i$ is the molar mass for the corresponding component
and
$\Phi(r)$ is the gravitational potential
at position $r$ \cite{hatsopoulos}.
Then, independent from the details of the gas radial distribution, 
$\log \rho_v \sim \log \rho_d$.
With somewhat more intricate calculation, 
the same conclusion can be drawn for a self-gravitating gas just in
hydrodynamic and not in full thermal equilibrium.
Consider a multi-component gas cloud in hydrodynamic equilibrium, then
for the density of each component we can write
\begin{equation}\label{dynamicequi}
\frac{d p_i(r)}{d r}+\rho_i(r) \frac{d \Phi(r)}{dr}=0,
\end{equation}
where $p_i(r)=\rho_i(r) T_i(r) / \mu_i$ is the partial pressure for component $i$.
After a simple manipulation we obtain
\begin{equation}\label{diff}
\frac{T_i(r)}{\mu_i} \frac{d \log \rho_i(r)}{dr}+\frac 1\mu_i\frac{dT_i(r)}{dr}+\frac{d\Phi(r)}{dr}=0
\end{equation}
and
\begin{equation}
\log\rho_i(r)=C-\mu_i \int \large[ d\log T_i(r) +\frac{d\Phi(r)} {T_i(r)} \large].
\end{equation}
This tells us that
$\log \rho_v \sim \log \rho_d \sim \int (d\log T + \frac 1T d\Phi )$
in case the components have temperatures
which are locally similar or equal.
Thus,  the above mentioned
log-log-linear correlations may contain important information about the
microscopic properties of the dark matter.
In this paper we would like to investigate in more details  the properties of
the mass profiles in the galaxy cluster CL0024.
Since in cosmology huge distances and small heat transfer rates seem to interfere
with the processes of thermal equilibration, thus reducing the likelihood
of thermal equilibrium, such study would be most certainly beneficial.

\begin{figure}
\centering
\begin{minipage}[c]{0.3\hsize}
\epsfig{file=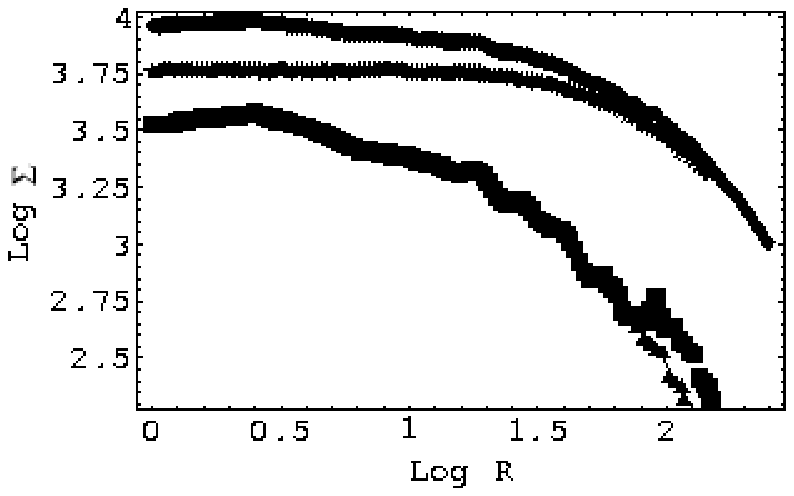,width=\hsize}
\end{minipage}
\hspace*{0.5cm}
\begin{minipage}[c]{0.3\hsize}
\epsfig{file=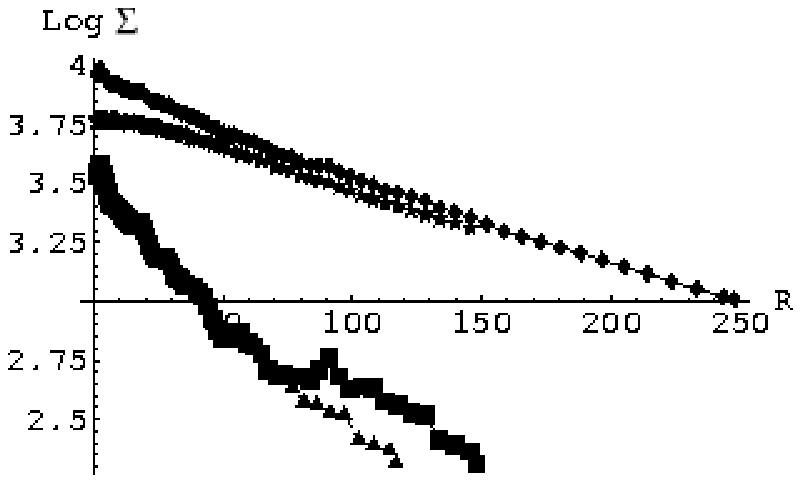,width=\hsize}
\end{minipage}
\hspace*{0.5cm}
\begin{minipage}[c]{0.3\hsize}
\epsfig{file=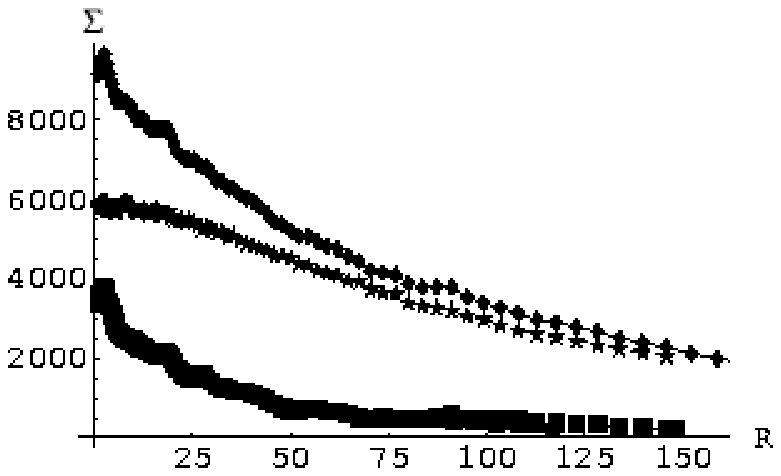,width=\hsize}
\end{minipage}
\caption{The projected mass density profiles from Tyson {\it et al.} plotted on different scales.
Up to down are total, dark and visible matter densities, respectively; triangles correspond
to distribution of visible matter corrected for the anomaly at $100 h^{-1}kpc$.}
\label{fig-1}
\end{figure}

In Ref.\cite{tyson} the projected radial density 
profiles for the total mass and the visible matter were originally presented on log-log scale
and mass profile for the dark matter could be straightforwardly deduced from these.
The total mass and the dark matter
form smooth density distribution 
monotonously decreasing with distance. 
Similar behavior is observed
for the visible matter
except for a flat segment at distance 
of about $100 h^{-1}kpc$ [see Fig.(\ref{fig-1})]. 
As can be seen from Fig.(\ref{fig-1}), at this distance
$\Sigma_v(R)$ is approximately constant for about $35 h^{-1}kpc$ while
all the other profiles continue to fall.
While we do not know the origin of such peculiar anomaly, we note that outside of this 
flat region the mass profile for visible matter has essentially the same behavior. In fact,
if we cut out this flat segment and shift the tail
of  the visible matter distribution to make a continuous curve, 
we would observe that the obtained profile is
almost a perfect straight line on $\log \Sigma - R$ scale
[data shown with triangles in Fig.(\ref{fig-1})].
Upon inspection, one notices that, actually,
on $\log \Sigma-R$ scale
all profiles
fall linearly with the distance,
which implies behavior
$\Sigma_i (R) \sim e^{-a_i R}$. 

\begin{figure}
\centering
\begin{minipage}[c]{0.3\hsize}
\epsfig{file=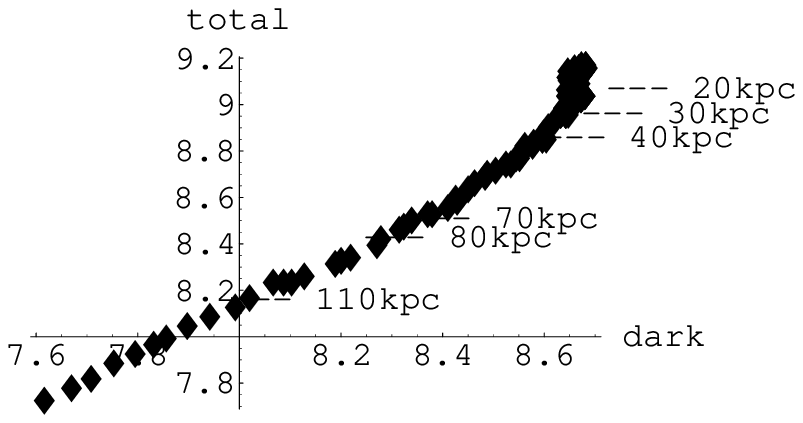,width=\hsize}
\end{minipage}
\hspace*{0.5cm}
\begin{minipage}[c]{0.3\hsize}
\epsfig{file=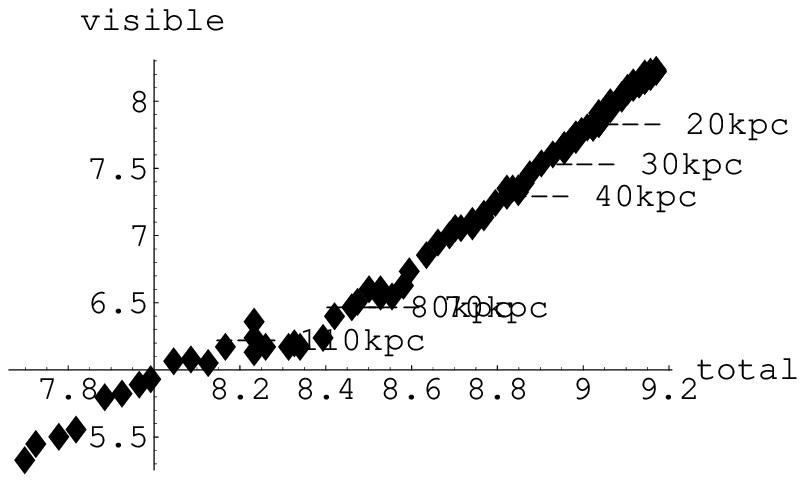,width=\hsize}
\end{minipage}
\hspace*{0.5cm}
\begin{minipage}[c]{0.3\hsize}
\epsfig{file=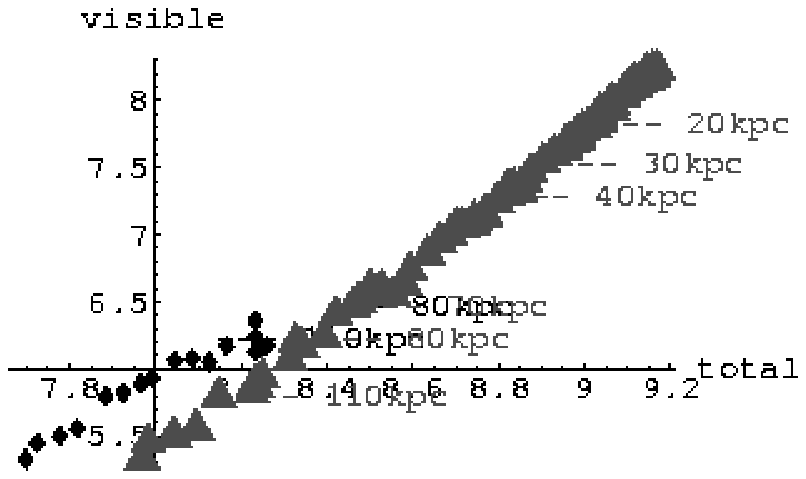,width=\hsize}
\end{minipage}
\caption{Linear correlations between mass profiles in galaxy cluster CL0024.
Shown are total and dark matter (left panel),
total and visible matter (center) and total and visible matter 
corrected for anomalous region (left panel, data shown with triangles).}
\label{fig-2}
\end{figure}

\begin{figure}
\centering
\begin{minipage}[c]{0.4\hsize}
\epsfig{file=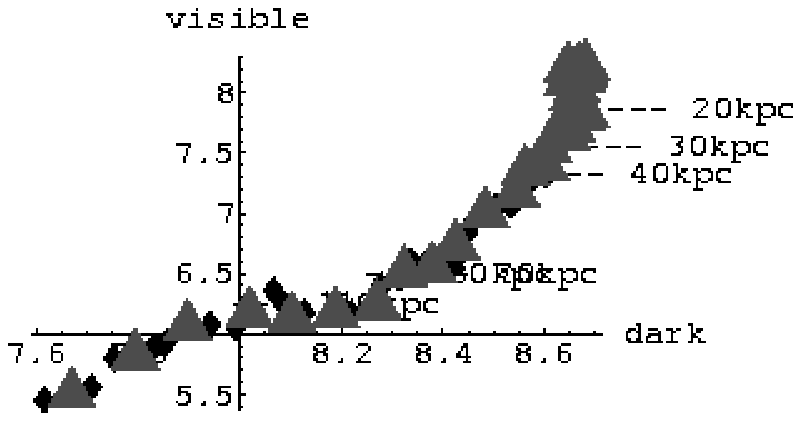,width=\hsize}
\end{minipage}
\hspace*{0.5cm}
\begin{minipage}[c]{0.4\hsize}
\epsfig{file=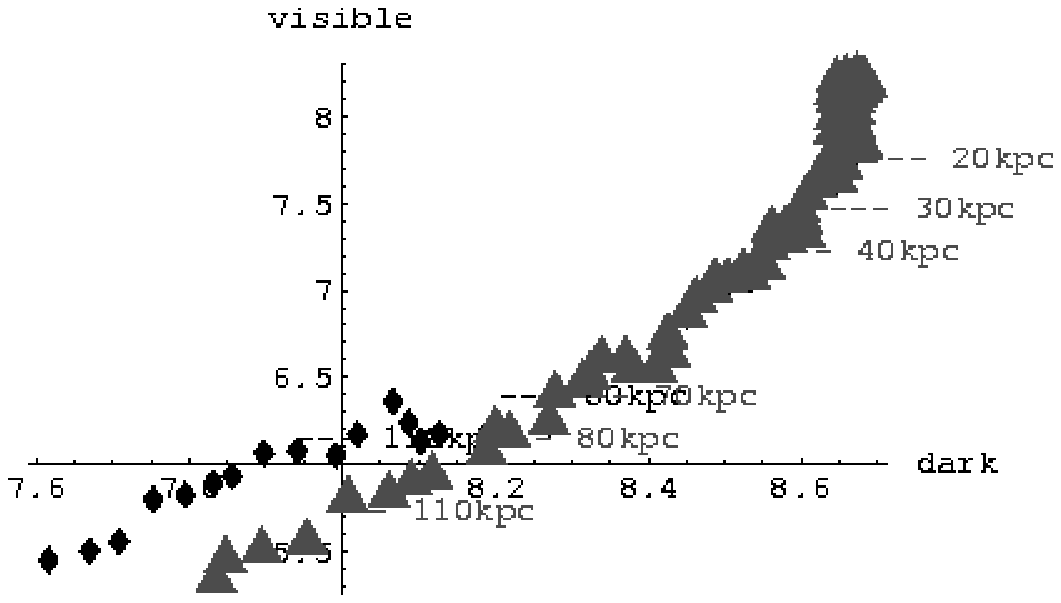,width=\hsize}
\end{minipage}
\caption{Linear correlation between log of dark and log of visible matter mass density.
Emphasized in the left panel 
is the anomalous behavior of the visible matter mass profile between $80 h^{-1}kpc$
and $110 h^{-1}kpc$.
Also shown is the log-log-linear correlation between dark and visible matter corrected
for anomalous region (data shown with triangles in the right panel).}
\label{fig-3}
\end{figure}

\begin{figure}
\centering
\epsfig{file=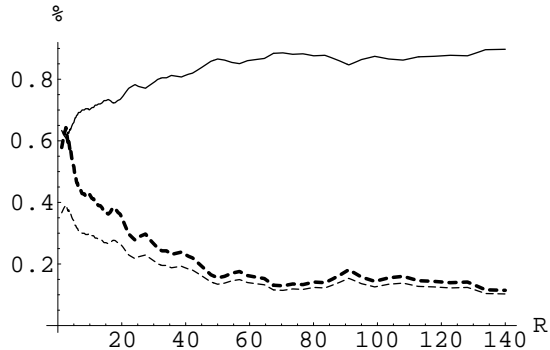, width=220pt}
\caption{Concentration by mass of the dark (solid line) and the visible (dashed line) matter in
CL0024 as function of distance $R$. Also shown is the ratio of these concentrations (thick
dashed line).}
\label{fig-X}
\end{figure}

As was already mentioned,
it can be noticed that mass distributions of dark and visible matter in CL0024 exhibit linear
correlation on log-log scale. Similar property can be observed also for other pairs
of mass profiles.
For example, if log of the total mass is plotted vs. log of the dark mass,
the linear correlation becomes very prominent. The same is true for the 
total and the visible matter mass densities. In this latter case the anomalous flat segment 
becomes more prominent; however, if the visible matter mass profile is corrected
for this region, the points on log-log scale form almost a perfect line [data shown
with triangles in the diagram in Fig.(\ref{fig-2})].
The relations between different mass profiles can be further analyzed by looking at different
segments of the diagrams in Fig.(\ref{fig-2}) and Fig.(\ref{fig-3}).
We found that the correlation coefficients 
tend to decrease somewhat if one moves away from the center of the cluster.
We obtained following correlation coefficients, where the error estimate
is related to their variation with distance:
$\kappa_{vt} \sim 2.45^{ + 0.1 }_{- 0.45}$,  
$\kappa_{td} \sim 1.5^{+0.35}_{-0.35}$,
$\kappa_{vd} \sim 3.6^{+0.8}_{-0.65}$, and
\begin{equation}
\label{correlation}
\log \Sigma_i \approx a_{ij} + \kappa_{ij} \log \Sigma_j.
\end{equation}
We also note that all three coefficients are consistent with each other and imply the 
concentration by mass of the dark matter about 80\%.
For example, from thermodynamic interpretation of 
$\kappa_{ij}\approx ( \mu_i /\mu_j)(T_j / T_i) $ we would expect
$\kappa_{vt}\kappa_{td}\approx \kappa_{vd}$ and this is indeed the case.
Furthermore, we can write for the total mass density
\begin{equation}
M(r) = m_d(r) + m_v(r) \approx a_d e^{-\mu_d \Phi (r)/ T} +  a_v e^{-\mu_v \Phi(r)/T},
\end{equation}
where $m_{d (v)}$ is density of dark (visible) matter and $\mu_{d (v)}$ is the molar mass
of dark (visible) matter.
Then, for variation $\delta M(r)$ we obtain
\begin{equation}
\begin{array}{c}
\delta M(r)\approx m_d(r) (-\mu_d \delta\Phi / T) + m_v(r) (-\mu_v \delta\Phi / T) \\
\approx - M(r) \large( \frac{m_d(r)}{M(r)} \mu_d + \frac{m_v(r)}{M(r)} \mu_v \large) \delta\Phi / T \\
\approx M(r) (-\mu_{tot} \delta\Phi / T),
\end{array}
\end{equation}
where 
\begin{equation}\label{totmolar}
\mu_{tot}=\frac{m_d(r)}{M(r)} \mu_d + \frac{m_v(r)}{M(r)} \mu_v 
\end{equation}
 is effective total molar mass. 
By dividing Eq.(\ref{totmolar})
with $\mu_v$ or $\mu_d$ and taking into account that,
for example, $\mu_{tot}/\mu_d = \kappa_{td}$,
one can immediately verify that all of the above-mentioned coefficients 
$\kappa$ satisfy Eq.(\ref{totmolar})
with the dark matter concentration by mass $m_d/M\approx 0.8$, consistent with 
direct observation from the data by Tyson {\it et al.} [see Fig.(\ref{fig-X})]

The existence of log-log-linear correlation in CL0024  becomes even more striking
once one realizes that similar relation exists in spiral galaxies. 
In spiral galaxies the presence of dark matter reveals itself through the 
anomalous behavior of the Rotation Curves (RC) 
which for large distances do not fall as $1/\sqrt{R}$, according to the Kepler's law, 
but remain roughly flat. 
Such behavior had been analyzed in the literature and fit by
synthetic  model of Persic and Salucci \cite{persic} 
in which a number of RC where categorized by galaxy luminosity and
reproduced  with superior accuracy with a simple analytic model (Universal RC) .
In this model the 
distribution of visible matter in the galaxy is described
by exponential thin disk with surface density
\begin{equation}
I_v(r) \sim e^{-3.2 r/R_{opt}},
\end{equation}
which is known to fit well the surface brightness of galaxy disks,
and dark matter is represented with a spherical flat-core distribution.
We noticed that the URC 
for all luminosities can also be successfully
described by a spherical dark hallo with density 
\begin{equation}
\rho_d (r) \sim e^{-a_d r/R_{opt}},
\end{equation}
in which case $\log I_v(r)$ and $\log \rho_d(r)$ are linearly related
with $\log I_v(R) \approx \kappa_{vd} \log\rho_d(R)$.
Surprisingly, the coefficient of such relation 
varies very insignificantly with galaxy luminosity
and corresponds to $\kappa_{vd}^{spiral} \sim 2.6-3.8$. 
This is consistent with $\kappa_{vd}$ obtained
from the analysis of log-log-linear correlation between the mass profiles in
the galaxy cluster CL0024.
Considering that such log-log-linear correlation is, in fact, 
expected from thermo- or hydrodynamic equilibrium, the above observation
is most indicative of the presence of a sort of 
equilibrium between the dark and
the visible component in spiral galaxies and galaxy clusters.

Let us now concentrate on the volume densities of visible and dark matter
in CL0024.
Note that the log-log-linear correlation observed above was obtained for
the projected mass densities $\Sigma_i(R)$, which are the quantities actually
measured in the gravitational lensing. 
At the same time $\rho_i(R)$, appearing in Eq.(\ref{equi}), is the volume mass density 
related to $\Sigma_i(R)$ via relation
\begin{equation}
\label{Abel}
\Sigma(r,\theta)=\int dz \rho (r,\theta,z).
\end{equation}
The projected mass density in general does not uniquely specify the volume density. Only with 
additional constraints, e.g. the spherical symmetry, Eq.(\ref{Abel}) can be inverted;
in spherically symmetric case this procedure is known as inverse Abel transformation  \cite{Abel}.
As is shown elsewhere, if mass distribution $\rho(R)$ decreases fast enough,
the projected mass density can be approximately represented as 
\begin{equation}
\label{projected}
\Sigma (r) \approx h(r) \rho(r),
\end{equation}
where $h(r)$ is a slow varying ($h(r) \sim r^\alpha$) effective depth and $\rho(r)$ is
the density maximum  along the line of integration in Eq.(\ref{projected}) \cite{ji_new}. 
The correlation between  $\log \Sigma_i(R)$
then implies
\begin{equation}
\begin{array}{rl}
\log \Sigma_i(r) &\approx \alpha_i \log r + \log \rho_i(r) \sim  \\
&\kappa_{ij} \log \Sigma_j(r) \approx \kappa_{ij} (\alpha_j \log r + \log \rho_j(r))
\end{array}
\end{equation}
and, if $ \log \rho(r) $ varies quicker than $\log r$, we get linear correspondence
\begin{equation}
\log \rho_i(r) \approx  \kappa_{ij}  \log \rho_j(r).
\end{equation}
Since we have observed before that the mass profiles $\Sigma_i(R)$
in the galaxy cluster CL0024
decrease exponentially with distance,
we do expect that linear correlation will be present
between $\rho$'s as well.
We can estimate the error introduced in our treatment because of the use of 
projected mass density $\Sigma(R)$
in place of the volume density $\rho(R)$ by inverting directly
the experimental profiles $\Sigma(R)$.

\begin{figure}
\centering
\epsfig{file=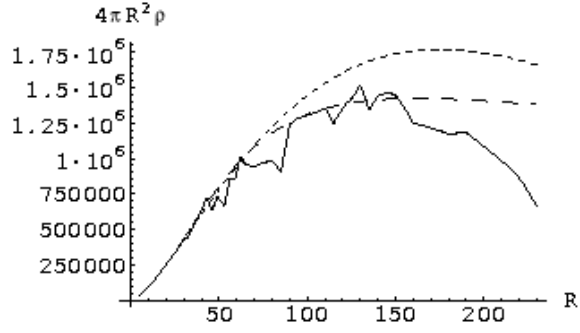,width=220pt}
\caption{Model dependence in volume density profiles 
obtained by inverse Abel transformation.
Solid line is extrapolation $\Sigma(R>R_{max})=0$, doted line is exponential fit and
dashed line is power-law fit. }
\label{fig-4}
\end{figure}

\begin{figure}
\centering
\begin{minipage}[c]{0.3\hsize}
\epsfig{file=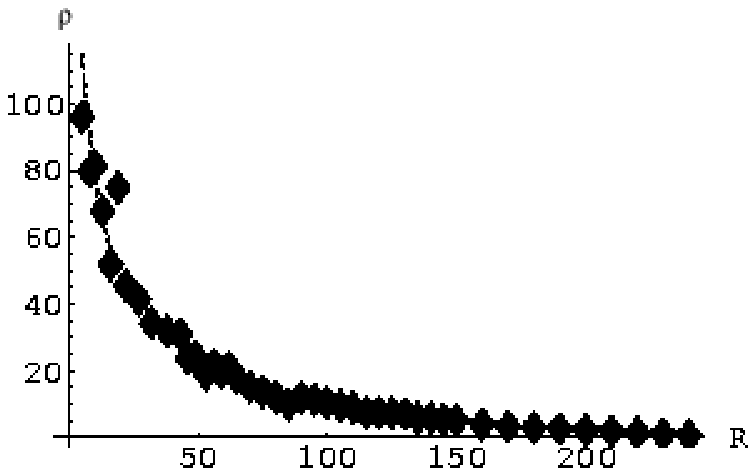,width=\hsize}
\end{minipage}
\hspace*{0.5cm}
\begin{minipage}[c]{0.3\hsize}
\epsfig{file=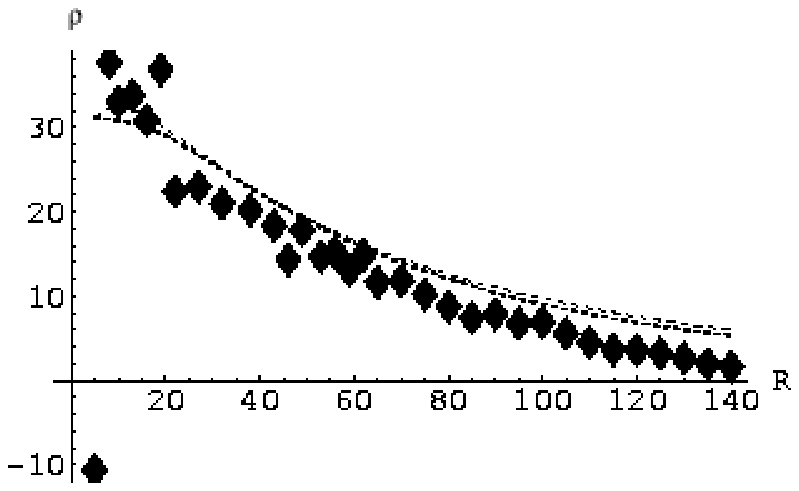,width=\hsize}
\end{minipage}
\hspace*{0.5cm}
\begin{minipage}[c]{0.3\hsize}
\epsfig{file=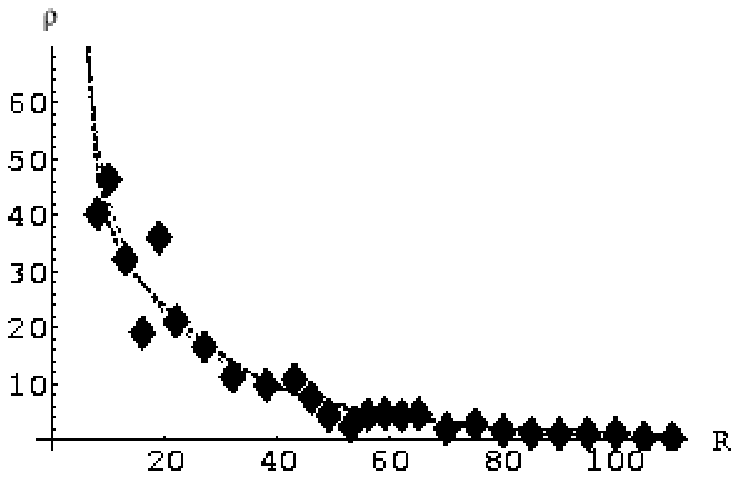,width=\hsize}
\end{minipage}
\caption{Radial volume density profiles obtained by inverse Abel transformation for
total mass, dark matter and visible matter (left, center, right respectively).
Points correspond to $\Sigma(R>R_{max})=0$ extrapolation, dashed line corresponds to exponential fit
and thick dotted line corresponds to power-law fit.}
\label{fig-5}
\end{figure}

Unfortunately, the inverse Abel transformation requires knowledge of $\Sigma(R)$
for arbitrary large $R$ while experimentally we only know a limited segment
$R\leq R_{max}\approx 120 h^{-1}kpc$ for visible 
and dark components and $R\leq R_{max}\approx 220 h^{-1}kpc$ for
total mass profile.
Certain strategy must be used to extrapolate the experimental profiles to 
$R>R_{max}$. In this study we adopt three main strategies:
cutting the integration in the inverse Abel transformation to $R<R_{max}$,
thus taking $\Sigma(R>R_{max})=0$;
fitting profiles with a sum of exponentials 
$\Sigma(R)=\sum a_i e^{-R/R_i}$ or fitting profiles with a sum of power-law functions
$\Sigma(R)=\sum a_i (1+R/R_i)^{b_i}$.
As should be expected, in every approach
obtained radial profiles are similar for $R$ small; 
however, at larger distances
a significant model-dependence becomes increasingly apparent.
In fact, obtained mass profiles are only model-independent
for the central region of the galaxy cluster out 
to about $R_{max}/2$ [see Fig.(\ref{fig-4})]. 
At the same time the central part of these profiles is most
sensitive to the noise in 
$\Sigma(R)$, so that we inevitably obtain
large errors and noise in $\rho(R)$ practically for all values of $R$ [see Fig.(\ref{fig-5})].
\begin{figure}
\centering
\begin{minipage}[c]{0.4\hsize}
\epsfig{file=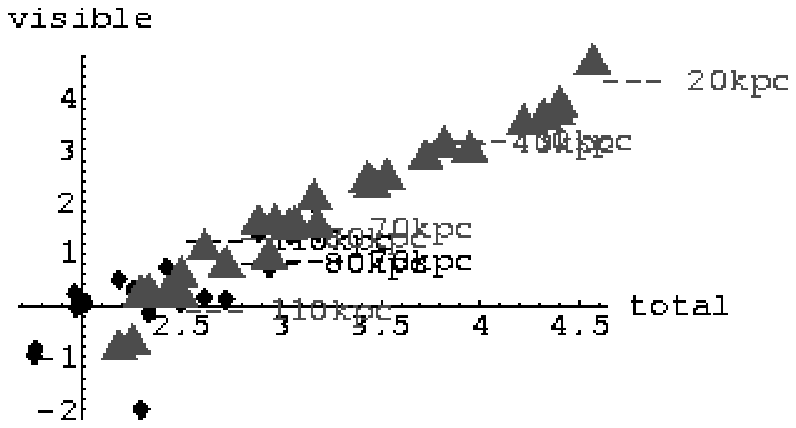,width=\hsize}
\end{minipage}
\hspace*{0.5cm}
\begin{minipage}[c]{0.4\hsize}
\epsfig{file=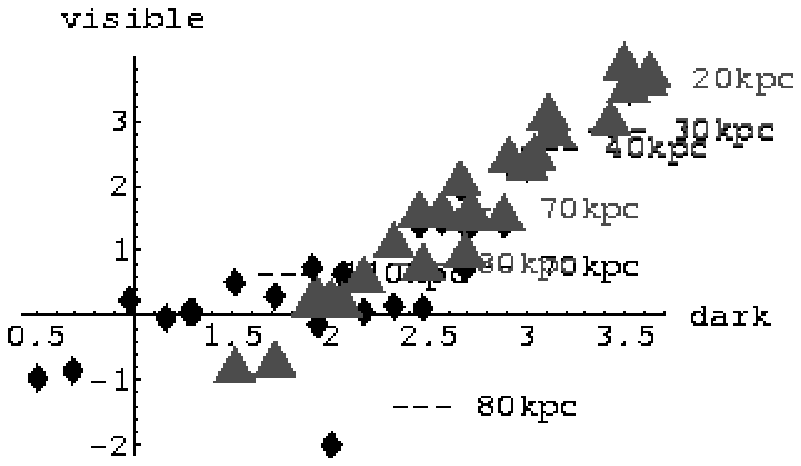,width=\hsize}
\end{minipage}
\caption{Inverted radial mass densities compared on log-log scale. As before, triangles denote the visible 
matter
distribution corrected for the anomalous region.
}
\label{fig-6}
\end{figure}

While data-points after inverse Abel transformation are more noisy, 
the linear correlation still can 
be seen on the log-log scale between any two mass profiles [see Fig.(\ref{fig-6})]. 
The parameters of this correlation are somewhat dependent on the
way the original $\Sigma$-distributions were extrapolated, but such
variation is well within acceptable bounds.
Taking into account this variation, for the parameters of
log-log-linear correlation between volume density profiles
we obtain
$\kappa_{vt}\sim 1.2-2.1$;
$\kappa_{td}\sim 1.0-2.0$ and
$\kappa_{vd}\sim 2.1-3.4$.
As we can see, the correlation coefficients are close and generally
somewhat less from those derived from the projected mass densities;
the average error introduced using the projected mass density $\Sigma(R)$
in place of volume mass density $\rho(R)$ is of the order of 30\%.

To understand the origin and magnitude of this error, let's consider effective depth  
defined by $\Sigma(R)=h(R) \rho(R)$. Then
\begin{equation}
\log \Sigma(R) = \log h(R) + \log \rho(R),
\end{equation}
and if
$\rho(R)$ falls off rapidly with $R$, one expects 
that $\log h(R)$ will vary relatively slowly with distance and thus the linear 
correlation between $\log \Sigma(R)$ would imply a linear correlation
between $\log \rho(R)$.
Nonetheless, variation of $\log h(R)$ would affect the estimate for $\kappa$
and
introduce an error of the order of $\Delta\log h(R) / \Delta \log \rho(R)$.
For example,
for the profile of dark matter in the region of interest
$\Delta \log h\approx 0.5$ and
$\Delta \log \rho\approx 1.5$. As is expected, 
the bias
introduced by this in our estimate of $\kappa_{vd}$ is of the order of 
$\Delta \log h / \Delta \log \rho \approx 30\%$.

\begin{figure}
\centering
\begin{minipage}[c]{0.4\hsize}
\epsfig{file=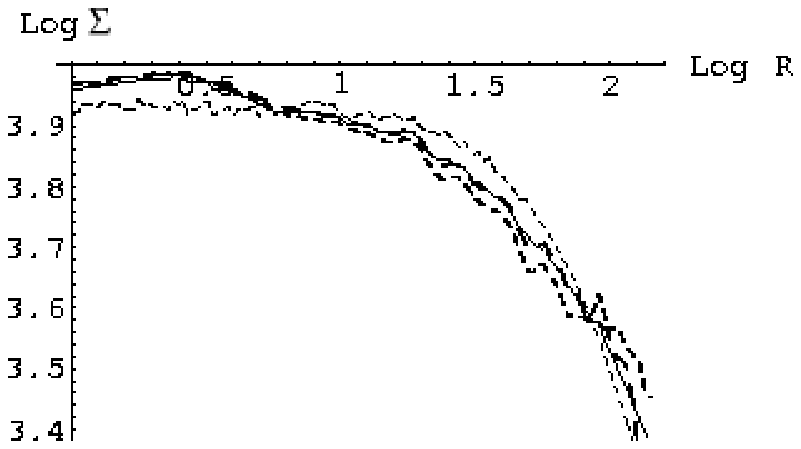,width=\hsize}
\end{minipage}
\hspace*{0.5cm}
\begin{minipage}[c]{0.4\hsize}
\epsfig{file=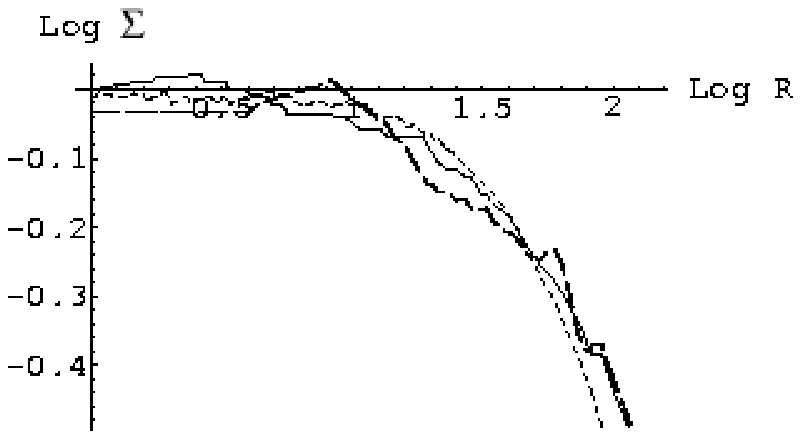,width=\hsize}
\end{minipage}
\caption{Alignment of the projected mass profiles in CL0024 by log-scaling (left)
and distance-scaling (right).}
\label{fig-7}
\end{figure}
Let us now turn our attention to
the alignment properties of the mass profiles in CL0024.
In Ref.\cite{ji} it was noticed that all mass profiles
can be aligned very well by either log-scaling [$\log \Sigma_i(r) \sim \kappa_{ij} \log \Sigma_j(r)$] or 
distance-scaling 
[$\log \Sigma_i(r) \sim \log \Sigma_j (r e^{-z_{ij}})]$, as can be seen from Fig.(\ref{fig-7}).
For the parameters of such alignment we find that the dark matter profile
can be aligned with the total mass profile by rescaling with $\kappa_{td} \sim 1.25-1.45$,
the mass profile for visible matter can be aligned with that for total mass with
$\kappa_{vt} \sim 2.0-2.5$ and the visible mass density profile can
be aligned with that for dark matter with
$\kappa_{vd} \sim 2.5-4$. These values are similar to those obtained from
log-log-linear correlation. In fact, it becomes obvious that log-scaling alignment
is directly related to the existence of
log-log-linear correlation and vice-versa.

\begin{figure}
\centering
\begin{minipage}[c]{0.4\hsize}
\epsfig{file=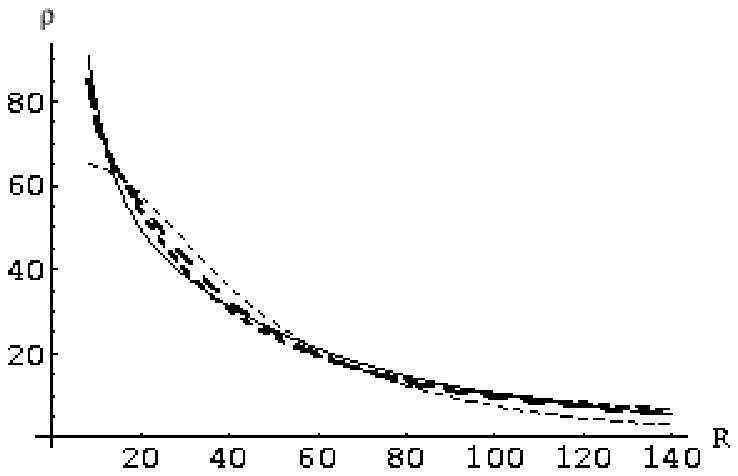,width=\hsize}
\end{minipage}
\hspace*{0.5cm}
\begin{minipage}[c]{0.4\hsize}
\epsfig{file=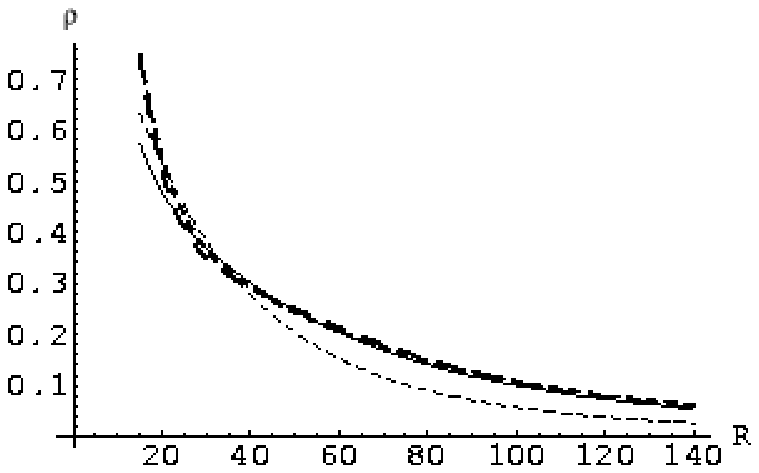,width=\hsize}
\end{minipage}
\caption{Alignment of the volume mass density profiles in CL0024 by log-scaling (left)
and distance-scaling (right).}
\label{fig-8}
\end{figure}

Similarly, all profiles can be aligned by distance-rescaling 
$\Sigma(r)\rightarrow \Sigma(r\cdot e^{-z})$.
We found that the best fit parameters are
$z_{vd}\approx -0.15- -0.25$ for visible-to-dark 
alignment, $z_{vt}\approx 0.45- 0.5$ for visible-to-total alignment
and $z_{dt}\approx 0.6-0.9$ for dark-to-total alignment.
The implication of these alignment properties is less obvious;
however, it  can be shown that together with log-scaling  they imply
\begin{equation}
\log \Sigma_i (r) = -a 10^{\gamma ( \log(r) - z_i) } + b_i=-a (\frac r{r_i})^\gamma + b_i.
\end{equation}
From the parameters $\kappa$ and $z$ listed above we find
$\gamma \approx 0.8-1.0$. 
This conclusion is consistent with our earlier observation that
the projected mass density in galaxy cluster CL0024 fall
exponentially with distance;
\begin{equation}
\Sigma_i(r) = B_i e^{-a (r/r_i)^\gamma}, \gamma \approx 1.
\end{equation}
If this result to be interpreted thermodynamically,
$\log \rho(R) \sim \Phi(R)$,
that would imply that the gravitational potential is rising in the region of interest almost
linearly.
Let us note that the same alignment properties are observed
for the volume mass densities as well.
All three volume density profiles can be aligned very well with the above-mentioned
scaling transformations [see Fig.(\ref{fig-8})].
The parameters of these alignments  are
$\kappa_{td}\approx 1.5-1.8 $,
$\kappa_{vt}\approx 1.7-1.9$,
$\kappa_{vd}\approx 2.8-3.4$
and
$z_{vd}\approx -0.35- -0.5$,
$z_{vt}\approx 0.4- 0.7$ and
$z_{dt}\approx 0.6- 0.8$,
similar to those obtained from  $\Sigma(R)$.

\section{Mass profiles and the thermal state of matter}
\label{secIII}

As was mentioned above, the log-log-linear correlation between mass profiles 
of the galaxy cluster CL0024 is a very remarkable property, especially given that 
a similar observation holds for a variety of spiral galaxies and that
such correlation is actually expected in case of
thermodynamic or hydrodynamic equilibrium. 
The most  tempting interpretation of such correlation
is through the thermo- (hydro-) dynamic equilibrium,
in which case the $\kappa_{vd}\approx 3.0$ should be related to the ratio 
of the molar masses
for visible and dark matter \cite{ji}.
This assertion, however, is difficult to understand theoretically from what
is known or assumed today about the properties of the dark matter \cite{ji_new}.
While there exist no direct data indicative of
the thermal state of the dark matter, the commonly accepted notion is that 
the dark matter is extremely weakly interacting and, thus, may well be thermally isolated
from the visible matter.
Here we would like to see if one can obtain any information about the
thermal state of the matter in 
galaxy cluster CL0024 from the actual mass profiles provided by Tyson {\it et al.}.

According to Eq.(\ref{equi}) and Eq.(\ref{dynamicequi}), 
the temperature of matter may be estimated from
$\rho(R)$ using
\begin{equation}
\label{bound}
\log\rho(R)-\log \rho(0)= \frac \mu T ( \Phi(0)-\Phi(R) )
\end{equation}
or from solving the differential Eq.(\ref{diff}) with respect to $T(R)$ with $\rho(R)$ known.
In the latter approach one needs to overcome the difficulty that the result of 
integration in Eq.(\ref{diff}) depends on unknown integration constant, for example
$T(R_{max})$. This problem may be partially avoided by 
integrating  Eq.(\ref{diff})
from large distance toward smaller $r$ to reduce dependence on the unknown boundary condition
$T(R_{max})$ or by using $T(R_{max})$ close to that
obtained from  Eq.(\ref{bound}).
In either case it can be observed that in the region of interest the 
dependence on boundary condition $T(R_{max})$ is weak.

\begin{figure}
\centering
\begin{minipage}[c]{0.4\hsize}
\epsfig{file=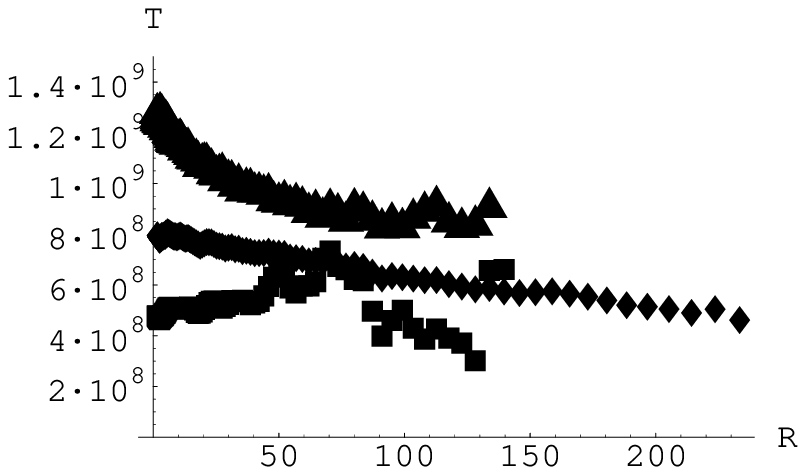,width=\hsize}
\end{minipage}
\hspace*{0.5cm}
\begin{minipage}[c]{0.4\hsize}
\epsfig{file=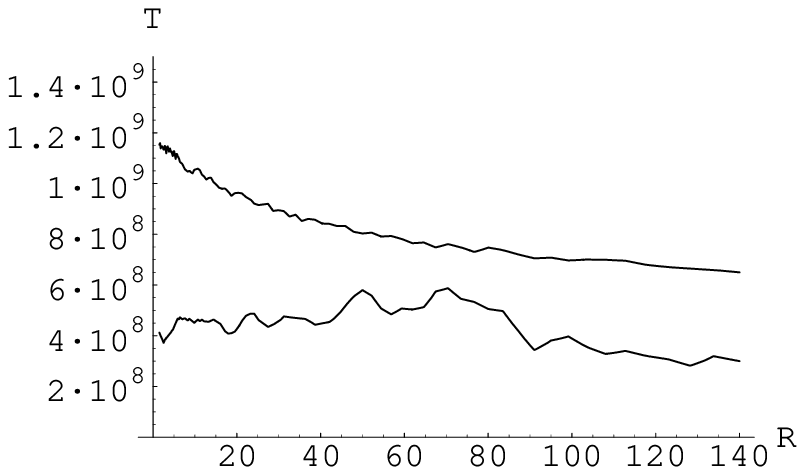,width=\hsize}
\end{minipage}
\caption{Temperature profile extracted from $\Sigma(R)$.
In the left panel shown is the temperature extracted from Eq.(\ref{bound})
and in the right panel shown is the temperature extracted from Eq.(\ref{diff}).
Triangles (higher line) correspond to dark matter temperature profile, boxes (lower line)
correspond to visible matter and diamonds correspond to effective temperature extracted from
the total mass distribution.}
\end{figure}
If the above idea is applied to the galaxy cluster CL0024 with the assumption
$\rho(R) \sim \Sigma (R)$, as above, 
the temperature extracted in this way indicates that the dark matter has smooth
temperature profile cooling outward, while the visible matter
is mostly isothermal with temperature somewhat decreasing in the center of the galaxy cluster. 
The estimate for the temperature of the visible matter $T_v\approx 5\cdot 10^8 K$ and
its decrease toward the center of the cluster is consistent 
with our understanding of X-ray clusters and 
the effects of radiative cooling in their centers \cite{sarazin}.
Results obtained in this way, however, are only qualitatively reliable.

Since $\log \rho(R) \approx -\log h(R) + \log \Sigma(R)$, in the case
when the temperature profile
$T(R)$ is almost isothermal even small correction $\log h(R)$ may be
significant. 
Generally, we expect
the "effective" depth $\log h(r)$ to increase the temperature of the
distribution $\Sigma(R)$ in the center of the cluster 
relative to the temperature associated with $\rho(R)$.
Obviously, that may have a dramatic  effect on our results making, for example,
the temperature profile for the dark matter isothermal or breaking isothermality
of the visible matter.

\begin{figure}
\centering
\epsfig{file=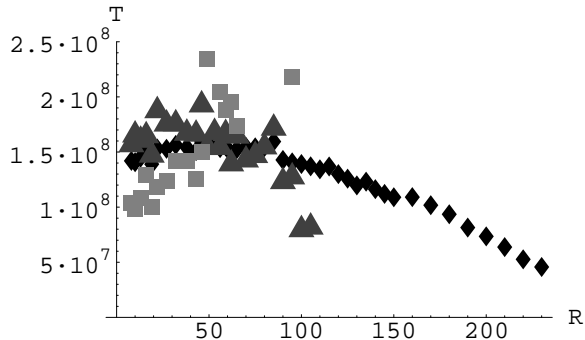,width=220pt}
\caption{Temperature profiles from volume densities deprojected
with $\Sigma(R>R_{max})=0$.
Triangles denote  dark matter temperature, boxes correspond to visible matter 
and diamonds correspond to effective temperature of the total mass profile.}
\label{fig-1a}
\end{figure}

Thus, we do need to consider the above approach applied directly
to the volume density profiles $\rho(R)$. 
As we have noticed before, the data points after
this transformation are significantly more noisy. 
This problem is even more serious here than in the case of log-log-linear correlations
since the resultant $T(R)$-profile is particularly sensitive to the 
noise in $\rho(R)$.
In fact, the
volume density profile obtained with $\Sigma(R>R_{max})=0$ is so noisy that hardly any conclusion can be 
obtained from it at all. As can be seen in Fig. (\ref{fig-1a}), the only conclusion
we can draw is that the temperature of the visible matter
is definitely decreasing toward the center of the cluster and that
the "effective" temperature obtained from the total distribution
is decreasing in the outer regions of the galaxy cluster.

Better results can be obtained when the mass profiles $\Sigma(R)$ are
smoothened and extrapolated with
a linear combination of exponential or power-law functions.
For both fits we obtain a similar result that the visible matter is cooling significantly in
the center of the cluster, consistent with the strong radiative cooling in X-ray clusters
mentioned above. At the same time for the dark matter
we obtain the distribution close to "isothermal" which typically cools very slightly in the center.
It becomes obvious, therefore, that the local thermal equilibrium between dark and visible
matter is significantly broken, at least, in the center of the galaxy cluster.
Unfortunately, 
while these qualitative conclusions are similar for both exponential and power-law extrapolations,
the quantitative details differ dramatically between the two, especially for the outer regions
of the cluster. 
The integration of Eq.(\ref{diff}) using exponential
extrapolation
produces temperature
distributions for the dark and the visible components with almost constant ratio 
$\kappa_{vd} \approx 1.8$.
This ratio gets larger toward the center of the galaxy cluster
as the visible matter cools rapidly and the dark matter
fails to follow the suit. The result of analysis for this case,
thus, supports
the hypothesis of overall thermal equilibrium. 
It implies that, although the ratios $T/\mu$ for
visible and dark matter vary with distance, the ratio 
$\kappa_{vd}=(T_d/T_v)( \mu_v/\mu_d)$ 
remains constant for a wide range of values of $R$.
The thermal equilibrium is only broken in the central part of the cluster 
where the rapid radiative cooling
of the visible component is essential.

However, the conclusion obtained using the power-law extrapolation is
exactly the opposite. 
In this approach
we do find that at large distances the ratios $T/\mu$ for the visible and the dark matter is
practically the same with $\kappa_{vd} \approx 1$.
This would be consistent with our understanding of the
cluster formation. According to our current understanding, the primary heat source
in galaxy clusters is gravitational heating.
Then, as originally cold gas collapses thereby forming galaxy cluster,
the gas molecules acquire kinetic energy $\Delta E_i \approx \mu_i \Delta \Phi$,
where $\Delta \Phi$ is the change in gravitational potential
during the collapse.
Later this kinetic energy is transfered to thermal energy of the matter
inside the cluster, so that $T_i \approx \mu_i \Delta\Phi$ and
thus $T_i /\mu_i \approx \Delta \Phi$ is the same for all kinds of
the particles in the gas cloud.
We expect therefore the mass distributions for different components
to be similar and $\kappa_{vd}\approx 1$, consistent with the result obtained with
power-law extrapolation.
$\kappa$ is not equal to 1 only
in the center of the cluster where the visible distribution cools rapidly
while the dark matter distribution remains almost isothermal. 
The temperature of the visible matter in both approaches comes out
$T_v\approx 10^8 K$.
\begin{figure}
\centering
\begin{minipage}[c]{0.3\hsize}
\epsfig{file=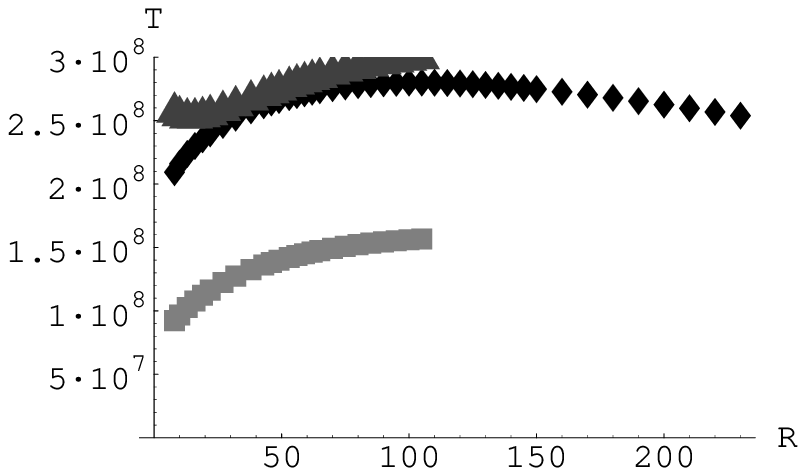,width=\hsize}
\end{minipage}
\hspace*{0.5cm}
\begin{minipage}[c]{0.3\hsize}
\epsfig{file=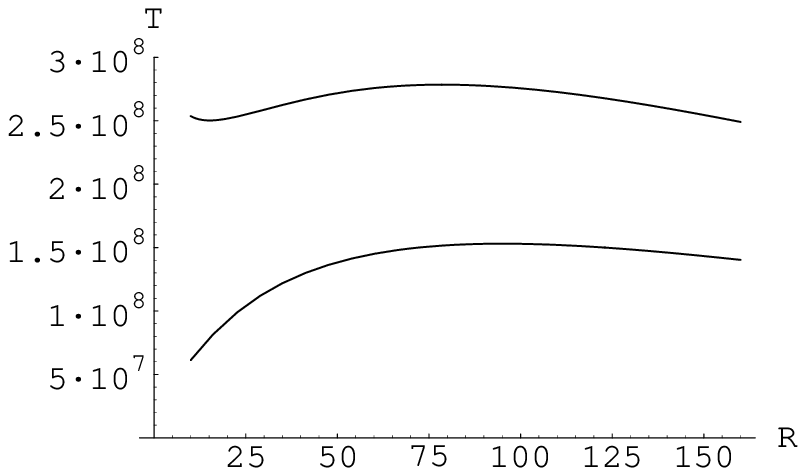,width=\hsize}
\end{minipage}
\hspace*{0.5cm}
\begin{minipage}[c]{0.3\hsize}
\epsfig{file=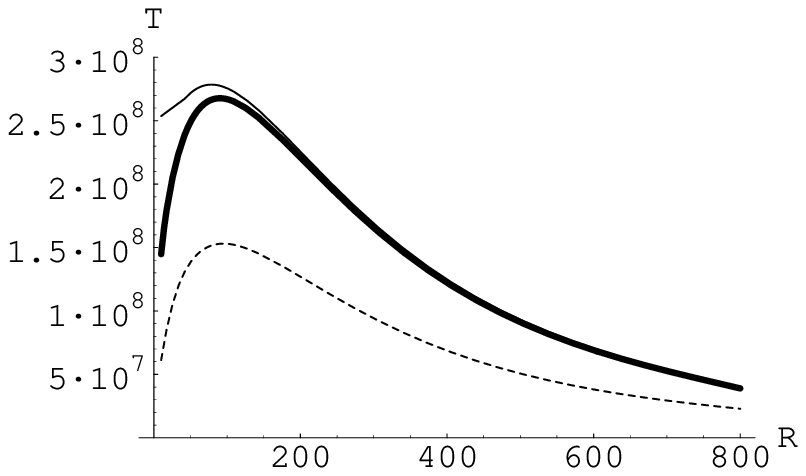,width=\hsize}
\end{minipage}
\caption{Temperature distributions for exponentially extrapolated $\Sigma(R)$.
Shown are $T(R)$ given by Eq.(\ref{bound})  (left panel), obtained from
 Eq.(\ref{diff}) (central panel) and extrapolation to large distances (right panel).
Higher curve corresponds to dark matter, while lower curve corresponds to
visible matter and the middle curve corresponds to "effective"
temperature obtained  for total mass profile.}
\end{figure}
\begin{figure}
\centering
\begin{minipage}[c]{0.3\hsize}
\epsfig{file=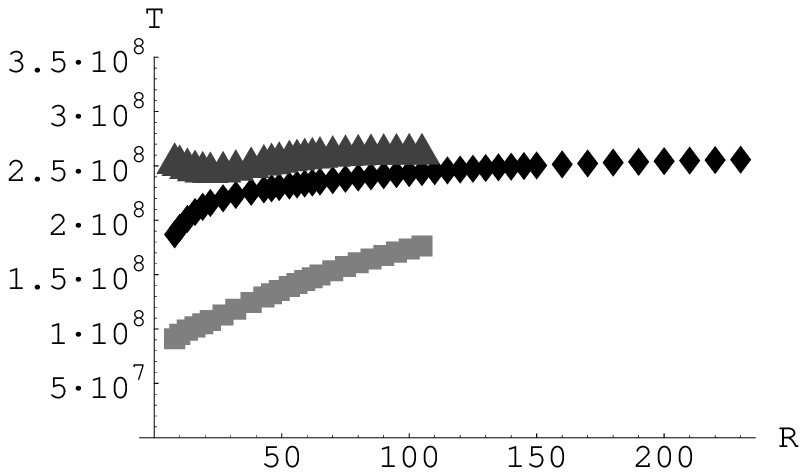,width=\hsize}
\end{minipage}
\hspace*{0.5cm}
\begin{minipage}[c]{0.3\hsize}
\epsfig{file=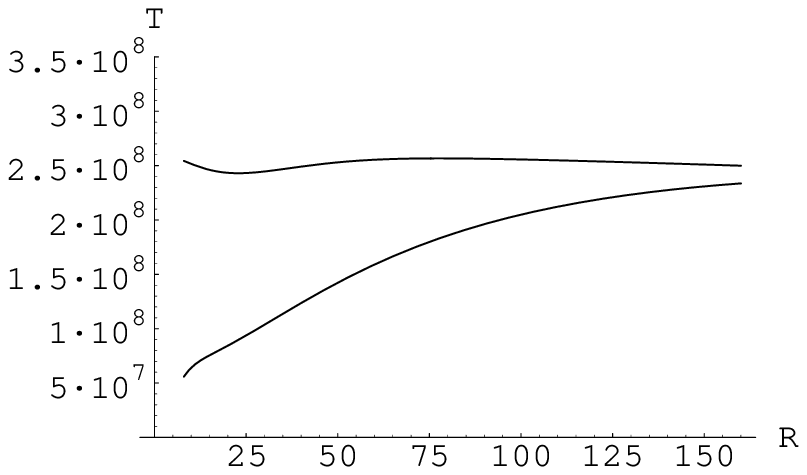,width=\hsize}
\end{minipage}
\hspace*{0.5cm}
\begin{minipage}[c]{0.3\hsize}
\epsfig{file=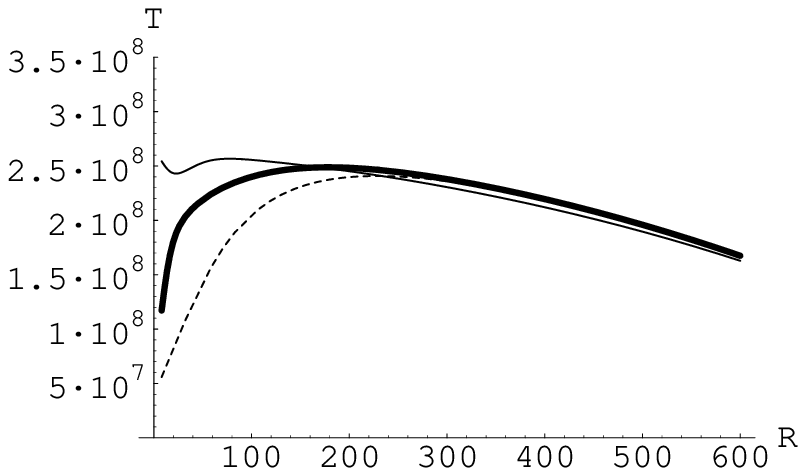,width=\hsize}
\end{minipage}
\caption{Temperature distributions for power-law extrapolated $\Sigma(R)$.
Shown are $T(R)$ given by Eq.(\ref{bound})  (left panel), obtained from
 Eq.(\ref{diff}) (central panel) and extrapolation to large distances (right panel).
Higher curve corresponds to dark matter, while lower curve corresponds to
visible matter and the middle curve corresponds to "effective"
temperature obtained  for total mass profile.}
\end{figure}

We must conclude, therefore, that the data on the distribution of mass in 
CL0024 are  not sufficient neither to support nor
to reject the hypothesis of thermal equilibrium. 
Only general conclusion can be drawn from this analysis.
In particular, the temperature extracted from the mass profile of the visible matter 
is of the order of $5\cdot 10^7-1.5\cdot 10^8 K$ and 
is consistent with X-ray  data on galaxy clusters.
The temperature of the visible component drops significantly in the center
of the cluster and
also falls slowly at large distances. 
The dark matter distribution appears to be close to isothermal with indications of slight
cooling in the center. 
We believe that such 
behavior may be indicative of nonvanishing thermal connection between 
the dark and the visible components 
in the cluster.
Still, the data are too uncertain to make this assertion any more definite.

The assumption of the local thermal equilibrium is most
certainly violated, at least in the central region, 
where the effect of radiative cooling
of the visible component is the strongest. In the other parts of the galaxy cluster, however,
we can neither confirm nor reject the hypothesis of thermal equilibrium as
conclusion depends strongly on the way to interpolate the available data to large values of $R$.

\section{Conclusion}
\label{conclusions}
In this work we extend the analysis of the visible matter, dark matter and total projected mass profiles
derived for the galaxy cluster CL0024 from strong gravitational lensing by Tyson {\it et al.}.
We observe that the linear correlation exists between each pair of mass profiles
on log-log scale. This linear correlation is preserved whether one uses the projected mass
profiles [$\Sigma(R)$] or the volume density profiles [$\rho(R)$]. 
The difference between the coefficients
of log-log-linear correlations obtained for $\Sigma(R)$ and $\rho(R)$ is small and is of the order
of 30\%.
We obtained following values for the parameters of log-log-linear correlation
between dark and visible mass distribution profiles
\begin{equation}
\begin{array}{l}
\kappa_{vd} \sim 2.9-4.4 $ (from projected mass profiles $\Sigma), \\
\kappa_{vd}\sim 2.1-3.4 $ (from volume density profiles $\rho), \\
\kappa_{vd}\sim 2.2-4.0 $ (from alignment properties for $\Sigma), \\
\kappa_{vd}\sim 2.8-3.3 $ (from alignment properties for $\rho).
\end{array}
\end{equation}
The correlation coefficients for other pairs of mass profiles were also listed in this paper.
We also analyzed in details the alignment properties of the mass profiles in the galaxy cluster.
We found that the log-scaling and the distance-scaling, mentioned earlier
in the literature, imply (and are consistent with) exponential behavior of the mass profiles in the region
of interest
\begin{equation}
\rho \sim e^{-a (r/r0)^\gamma}, $  $\gamma \approx 0.8-1.0.
\end{equation}

These properties of the mass profiles are striking, especially
since similar correlations are observed in the spiral galaxies with
$\kappa_{vd}\sim 2.6-3.8$. Given huge differences between the systems
where such correlations are observed, it appears that this cannot have
an accidental character but is related to actual properties of dark and visible matter.
Interestingly, such behavior of the mass profiles in self-gravitating system
is expected in the case of thermo- or hydrodynamic equilibrium. 
Whenever the thermal states of the components in such system are similar, one expects
\begin{equation}
\log \rho_v \sim \log \rho_d.
\end{equation}
It is therefore tempting to interpret the observed correlations as a sign of 
equilibrium between the dark and the visible components in the
galaxy cluster in which case the ratio $\kappa_{vd}$ corresponds
to the molar mass ratio of visible and dark matter
\begin{equation}
\frac{\mu_v}{\mu_d} \approx 2.1 - 4.4.
\end{equation}
This suggests the mass of the dark matter particle between $\mu_d \approx 200-1000 MeV$.
To our knowledge, there are no candidates within this mass range in
the current extensions of Standard Model. Massive neutrinos and axions
have experimental mass limits $<25$MeV for the 
heaviest $\tau$-neutrino \cite{hu} which is well below our range.
The most favorable candidates for non-baryonic dark matter, such as 
Weakly Interacting Massive Particle (WIMP) with the mass anywhere between 
10GeV and 1TeV and the SUSY lightest particle, {\it e.g.} neutralino, with 
the mass above 30GeV \cite{khalil}, are far above the range obtained
in our study. 
In fact, the closeness of $\mu_d$ to QCD-energy scale $\Lambda_{QCD}$ may indicate
that dark matter is ultimately related to QCD phenomena,
such as quark and gluon condensations.
Whether this is true and what kind of connection may exist between the dark matter and
QCD is an interesting topic for further discussion.

The concept of thermo- (hydro-) dynamic equilibrium between dark and visible matter
encounters significant theoretical difficulties in trying to understand the possible
source of equilibrium between, supposedly, extremely weakly interacting dark and visible components.
For that reason we further analyzed what kind of information about the thermal properties
of the matter in the galaxy cluster CL0024 can be extracted from available measurement.
In our analysis we found that for visible matter the average temperature is of the order of
$T_v \approx 10^8 K$, consistent with what we know about the temperature
of the intergalactic gas in X-ray clusters.
We also found that the temperature of the visible matter drops significantly 
in the center of the cluster, also consistent with what we know about 
radiative cooling in X-ray clusters.
For the dark matter we obtained almost isothermal temperature profile
with signs of slight cooling in the center of the cluster. We believe that such
behavior may indicate existence of thermal connection between dark and visible matter.
At the same time we found that the available data are insufficient to either confirm or
reject hypothesis of thermal equilibrium as the conclusion depends strongly on the
way the data are smoothened and extrapolated at large distances. While in the central part of the
cluster the equilibrium is most certainly broken 
by rapid radiative cooling of the visible component,
in the other parts of the galaxy cluster one can find signs
consistent either with full thermal equilibrium or full thermal isolation between
the dark and the visible components - depending on the way the projected mass density
profiles are extrapolated.

We must conclude therefore that no reliable quantitative conclusion
about the thermal state of matter in the galaxy cluster CL0024 can
be reached on the basis of known mass density profiles.

The observed similarities in the relative behavior of dark and visible matter in spiral galaxies
and galaxy cluster CL0024 are very puzzling and deserve further investigations. 
While the precision of individual mass distribution measurements
in galaxy clusters may be insufficient to obtain certain 
conclusions about the origin of such behavior, the synthetic analysis
of few such studies may be helpful.
One needs to see if the log-log-linear correlation can be found
in other galaxy clusters and if its coefficients have similar values to
those we have observed here.

\end{document}